\documentclass[a4paper,english,12pt,reqno]{amsart}

\usepackage{amsmath, amsthm, amssymb}
\usepackage{a4, babel, graphicx, listings, fancyhdr, verbatim}
\usepackage{mathrsfs} 
\usepackage{anysize} 
\usepackage{natbib}
\usepackage[format = hang]{caption}                              

\usepackage{a4}
\usepackage[latin1]{inputenc}
\usepackage{amsmath, amsthm}
\usepackage{graphicx}
\usepackage{amssymb}
\usepackage{verbatim}
\usepackage{listings}
\usepackage{alltt}
\usepackage{babel}
\usepackage{mathrsfs}
\usepackage{ae}
\usepackage{natbib}
\usepackage{booktabs} 

\newtheoremstyle{theorem}
{10pt} 
{10pt} 
{\sl} 
{\parindent} 
{\bf} 
{. } 
{ } 
{} 

\theoremstyle{theorem}

\def\beq{\begin{eqnarray}}
\def\eeq{\end{eqnarray}}

\def\beqn{\begin{eqnarray*}}  
\def\eeqn{\end{eqnarray*}}

\def\dd{{\rm d}}
\def\N{{\rm N}}

\def\Pr{P}

\def\arr{\rightarrow}
\def\hatt{\widehat}

\def\sumin{\sum_{i=1}^n}

\def\eps{\varepsilon}
\def\half{\hbox{$1\over2$}}

\def\rootn{\sqrt{n}}
\def\data{{\rm data}}
\def\obs{{\rm obs}}
\def\midd{\,|\,}

\def\dell{\partial}

\def\prof{{\rm prof}}

\def\calD{{\mathcal D}}

\def\Bernstein{{Bernshte\u\i n}}

\def\pois{{\rm Pois}}

\def\cc{{\rm cc}}

\def\unif{{\rm unif}}
\def\dev{{\rm dev}}
\def\piv{{\rm piv}}

\numberwithin{equation}{section} 
\numberwithin{figure}{section}
\numberwithin{table}{section}
\usepackage{hyperref}

\title[CDs and related themes]{Confidence Distributions and Related Themes}
\date{June 2017}

\begin{document}


\maketitle

\centerline{\bf Nils Lid Hjort$^1$ and Tore Schweder$^2$}

\medskip 
\centerline{\bf $^1$Department of Mathematics
    and $^2$Department of Economics}
\smallskip
\centerline{\bf University of Oslo} 

\begin{abstract}
This is the guest editors' general introduction to 
a Special Issue of the Journal of Statistical Planning
and Inference, dedicated to confidence distributions 
and related themes. Confidence distributions (CDs) 
are distributions for parameters of interest, 
constructed via a statistical model after analysing 
the data. As such they serve the same purpose 
for the frequentist statisticians as 
the posterior distributions for the Bayesians. 
There have been several attempts in the literature 
to put up a clear theory for such confidence distributions,
from Fisher's fiducial inference and onwards. There
are certain obstacles and difficulties involved 
in these attempts, both conceptually and operationally, 
which have contributed to the CDs being slow 
in entering statistical mainstream. Recently
there is a renewed surge of interest in CDs 
and various related themes, however, reflected 
in both series of new methodological research, 
advanced applications to substantive sciences, 
and dissemination and communication via workshops
and conferences.  
The present special issue of the JSPI is a collection
of papers emanating from the {\it Inference With Confidence} 
workshop in Oslo, May 2015. Several of the papers
appearing here were first presented at that workshop.
The present collection includes however also 
new research papers from other scholars in the field. 
\end{abstract}

\keywords{
\noindent {\it Key words:} 
confidence curves, 
confidence distributions, 
focus parameters, 
likelihood, 
meta-analysis, 
probability}


The Journal of Statistical Planning and Inference 
decided in the autumn of 2015 to arrange for a Special Issue
on confidence distribution and related themes. 
After various efforts, by patient authors, referees, 
and colleagues, along with the customary revision processes,
this has resulted in the current 
collection of eleven journal articles: 
\begin{itemize}
\item[1] \citet{CunenHermansenHjort17}, 
   on CDs and confidence curves for change points, 
   with applications to mediaeval literature 
   and to fisheries sciences; 
\item[2] \citet{DeblasiSchweder17}, 
   on median bias corrections for fine-tuning CDs; 
\item[3] \citet{Gruenwald17}, 
   on safe probability, leading also to tools for predictions;   
\item[4] \citet{Hannigetal17}, 
   on fusion learning and inter-laboratory analyses;   
\item[5] \citet{Lewis17}, 
   on combining inferences, with application to climate statistics; 
\item[6] \citet{LindqvistTaraldsen17}, 
   on proper uses of improper distributions; 
\item[7] \citet{Martin17}, 
   on generalised inference models;  
\item[8] \citet{Schweder17}, 
   with an essay on epistemic probability; 
\item[9] \citet{Shenetal17}, 
   on CDs for predictions, in different setups;  
\item[10] \citet{TaraldsenLindqvist17}, 
   on conditional fiducial models; and  
\item[11] \citet{VeroneseMelilli17}, 
   on CDs and their connections to objective Bayes.  
\end{itemize} 
These papers deal with theory and applications 
for distributional statistical inference, with 
CDs and fiducial distributions being the central concepts. 
Quite a few contributions also touch Bayesian angles 
and connections, however (\citet{CunenHermansenHjort17}, 
\citet{Gruenwald17}, \citet{Lewis17}, \citet{LindqvistTaraldsen17},
\citet{TaraldsenLindqvist17}, \citet{VeroneseMelilli17}). 
In the present general introduction to the Special Issue,
by the guest editors, efforts are made
both to explain to the broader statistical audience 
what confidence distributions (CDs) and confidence curves are; 
why and how they are steadily becoming more popular, 
in statistical theory and practice; and to briefly 
place the eleven papers in a broader context. 
In our article, which is by itself a gentle introduction 
to the general CD themes, we also attempt to point 
to aspects and issues and types of application 
not already contained in the review 
paper \citet{XieSingh13} and ensuing discussion. 


\section{The Holy Grail: frequentist posterior distributions}
\label{section:intro} 

Suppose data 
are analysed via some model, and that 
$\psi$ is a parameter of particular interest. 
Statisticians have many methods in their toolboxes
for conducting inference for $\psi$, such as
reaching a point estimate, assessing its precision, 
setting up tests, along with p-values when of relevance, 
finding confidence intervals, 
comparing the $\psi$ with other parameters from
other studies, etc. For the frequentist, constructing 
a distribution for $\psi$, given the available information, 
is more problematic, however, also conceptually.

Somehow it appears to be a strict Bayesian privilege 
to arrive at an appropriate posterior distribution,
say $p(\psi\midd\data)$ -- along with the associated 
difficulties of carrying out Bayesian work in the first place,
involving elicitation of prior distributions
and combining these with probability distributions of
a different kind. Working out a $p(\psi\midd\data)$
in the frequentist framework appears to clash 
with the basic premise that the parameter vector 
of the model is a fixed but unknown point in the parameter 
space. This has not stopped scholars from attempting
precisely such a feat, called the Holy Grail of 
parametric statistics by Brad Efron \citep{Efron10}. 
The earliest attempts were by none other than Sir Ronald Fisher, 
in a series of papers in the 1930ies  
\citep{Fisher30, Fisher32, Fisher33, Fisher35}. 
Certain obstacles and difficulties were found and pointed 
to by a number of critical scholars, however, and Fisher 
did not quite manage to defend his notion of a fiducial 
distribution for parameters. Indeed the fiducial ideas
have been referred to as `Fisher's biggest blunder'; 
see \citet[Ch.~6]{CLP16} 
for an account of the historical development,
and also \citet[this issue]{Gruenwald17}. 

There are however other and partly related notions of 
how to reach proper frequentist posterior distributions,
without priors, and the collective labels for a 
fair portion of these refined and modernised constructions 
are {\it confidence distributions} (CDs) and {\it confidence curves}.
There is a clear surge of interest in these methods
and in various related themes, regarding both 
theory and applications. This is witnessed in 
books and journal articles and by applied advanced work,
and is also reflected in high-level workshops and
conferences. The {\it BFF: Bayes, Frequentist, Fiducial}
series of conferences (also referred to as `Best Friends Forever') 
is reaching a steadily wider audience, 
with the current list being 
Shanghai (2014, 2015), Rutgers, New Jersey (2016), 
Harvard, Massachusetts (2017), 
Ann Arbor, Michigan (2018), and 
Duke and SAMSI, North Carolina (2019). 
There are also special invited sessions at major conferences, 
etc., dedicated to CDs and BFF themes. 
\citet{Efron98} speculates that Fisher's (alleged) 
biggest blunder might turn into a big hit for the 21st century;
see also \citet[Ch.~11]{EfronHastie16}.
   
The present special issue of the JSPI is dedicated
to such CDs and the growing list of related topics.  
The collection of papers and the ensuing organisation
of the special issue have grown out of one of 
these conferences, the {\it Inference With Confidence} 
workshop in Oslo in May 2015, organised by the 
the research group {\it FocuStat: Focus Driven Statistical 
Inference With Complex Data}. 
Some of the papers appearing 
in this issue were first presented as invited 
lectures at this workshop. We have also recruited 
contributions from other scholars in the field, however,
in an attempt to exhibit and see discussed 
a decent range of the more crucial dimensions 
of CDs and their increasing scope and usefulness,
in methodological and applied statistical work.  

``The three revolutions in parametric statistical inference 
are due to Laplace (1774), Gauss and Laplace (1809--1811) 
and Fisher (1922)'', is the clear opening statement 
in the two books \citet{Hald98, Hald06}. Somewhat boldly,
\citet[Preface]{CLP16} claim there is an ongoing fourth revolution
in statistics, at the start of the current millennium.
This fourth revolution has perhaps a less clear focus 
than the three drastic methodological changes Hald 
describes, and is arguably more about the {\it who}
and {\it what} than about the {\it how}, but we argue 
there that CDs and confidence curves have a natural place
in the world of statistical computation and communication,
also with Big Data. ``I wish I'd seen a confidence curve earlier'',
as tweeted J.M.~White, who manages a branch 
of Facebook's Core Data Science team, in April 2017. 
We should also make clear that
there by necessity are several approaches (partly 
related and partly competing) to the alleged Holy Grail
of reaching posteriors without priors. In addition
to the CD theory expounded in 
\citet{SchwederHjort02, SchwederHjort03, CLP16, XieSingh13}, 
with roots all the way back to Fisher in the 1930ies, 
there is generalised fiducial inference, 
see \citet{Hannigetal16} and \citet[this issue]{Hannigetal17}, 
along with \citet[this issue]{LindqvistTaraldsen17}
and \citet[this issue]{TaraldsenLindqvist17}; 
as well as the theory of inferential models, 
cf.~\citet{MartinLiu15} and \citet[this issue]{Martin17}. 
There is bound to be yet other hybrids and connections,
and some of these are touched upon in the present
collection of journal articles. 



\section{What are confidence distributions and confidence curves?}
\label{section:whatareCDs}

There are several ways in which to motivate,
define and construct such CDs, along with associated 
concepts and functions. Suppose the model for 
the data $y$ is governed by a parameter vector $\theta$, 
and that the interest parameter 
$\psi$ is a function $\psi(\theta)$ of the model parameter. 
A modern definition of a {\it confidence curve} for 
$\psi$, say $\cc(\psi,y)$, 
see \citet{SchwederHjort02, CLP16, XieSingh13}, 
is as follows. We write $Y$ for the random outcome
of the data generating mechanism and $y_\obs$ for the
actually observed data. At the true parameter point 
$\psi_0=\psi(\theta_0)$, the random variable $\cc(\psi_0,Y)$ 
should have a uniform distribution on the unit interval. Then 
\beq
\label{eq:ccbasic}
\Pr_{\theta_0}\{\cc(\psi_0,Y)\le\alpha\}=\alpha
   \quad \hbox{for all}\,\alpha. 
\eeq   
Thus confidence intervals, and more generally 
confidence regions, can be read off, at each desired level; 
the 90\% confidence region is $\{\psi\colon \cc(\psi,y_\obs)\le0.90\}$,
etc. When $\alpha$ tends to zero the confidence region
typically tends to a single point, say $\hatt\psi$,
an estimator of $\psi$. 
In regular cases the $\cc(\psi,y)$ is decreasing 
to the left of $\hatt\psi$ and increasing to the right, 
in which case the confidence curve $\cc(\psi,y)$
can be uniquely linked to a full confidence distribution 
$C(\psi,y)$, via  
\beq 
\label{eq:cases}
\cc(\psi,y)=|1-2\,C(\psi,y)|
   =\begin{cases}1-2\,C(\psi,y) &{\rm if\ }\psi\le\hatt\psi, \\
                 2\,C(\psi,y)-1 &{\rm if\ }\psi\ge\hatt\psi. \end{cases} 
\eeq 

The confidence name given to these post-data
summaries for focus parameters stems from the intimate
connection to the familiar confidence intervals. 
With $C(\psi,y)$ a CD, $[C^{-1}(0.05,y_\obs),C^{-1}(0.95,y_\obs)]$ 
becomes an equi-tailed 90\% confidence interval, etc.
Also, solving $\cc(\psi,y_\obs)=0.90$ yields two cut-off 
points for $\psi$, precisely those of the 90\% confidence interval. 
Correspondingly one may start with a given set
of nested confidence intervals, for all levels $\alpha$,
and convert these into, precisely, a CD. 


\section{General recipes}
\label{section:recipes}

Suppose a model with parameter vector $\theta$ is used 
for data $y$ and again that $\psi=\psi(\theta)$ 
is a focus parameter. If $\piv(\psi,y)$ is a function 
monotone increasing in $\psi$, with a distribution not 
depending on the underlying parameter, we term it a pivot. 
Thus $K(x)=\Pr_\theta\{\piv(\psi,Y)\le x\}$ 
does not depend on $\theta$, or on $\psi$, which implies that 
$$C(\psi,y)=K(\piv(\psi,y)) $$
is a CD. The classical construction of this type is 
that of \citet{Student08}, namely 
$$t={\mu-\bar y\over s/\rootn} $$
for a normal sample, with $\bar y$ and $s$ denoting the 
sample mean and empirical standard deviation. 
The ensuing CD for $\mu$ becomes 
$$C(\mu,\data)=F_\nu(\rootn(\mu-\bar y)/s), $$
with $F_\nu$ the cumulative distribution function of 
a $t$ distribution with the relevant degrees of freedom. 

In various classical setups for parametric models, there are
well-working large-sample approximations for the the behaviour
of estimators, deviance functions, etc., and these lead
to constructions of CDs and confidence curves. First, 
if $\hatt\psi$ is such that $\rootn(\hatt\psi-\psi)\arr_d\N(0,\tau^2)$,
and $\hatt\tau$ is a consistent estimator for the $\tau$ 
in question, then $\rootn(\hatt\psi-\psi)/\hatt\tau\arr_d\N(0,1)$.
Writing 
\beq
\label{eq:CDapprox1}  
C_n(\psi,\calD_n)=\Phi(\rootn(\psi-\hatt\psi)/\hatt\tau), 
\eeq
therefore, with $\calD_n$ the data available after $n$ observations, 
we have $C_n(\psi,\calD_n)\arr_d\unif$; in particular, the 
$C_n(\psi,\calD_n)$ is asymptotically a pivot in the above sense. 
Hence such a $C_n(\psi,\calD_{n,\obs})$ is a large-sample 
valid CD, allowing us to write 
\beq
\label{eq:CDapprox2}
\psi\midd\data\approx_d  \N(\hatt\psi,\hatt\tau^2/n), 
\eeq 
in the CD sense. This is akin to a Bayesian posterior 
distribution for $\psi$ (but without any notion of a prior
distribution involved). Also, the associated confidence curve,
asymptotically valid, is 
$$\cc(\psi,D_{n,\obs})=
   |1-2\,\Phi(\rootn(\psi-\hatt\psi_\obs)/\hatt\tau_\obs)|. $$

These first-order large-sample approximations 
(\ref{eq:CDapprox1})--(\ref{eq:CDapprox2}) are 
simple and useful but sometimes too coarse. A recipe 
that typically works better is the following. 
With $\ell_n(\theta)$ the log-likelihood function, let 
$\ell_{n,\prof}(\psi)=\max\{\ell_n(\theta)\colon\psi(\theta)=\psi\}$
be the profile, which we then turn into the 
deviance function 
\beq 
\label{eq:deviance}
\dev_n(\psi)=2\{\ell_{n,\prof}(\hatt\psi)-\ell_{n,\prof}(\psi)\}. 
\eeq 
By the Wilks theorem (see e.g.~\citet[Chs.~2-3]{CLP16}), 
under mild regularity conditions 
$\dev_n(\psi_0)\arr_d\chi^2_1$, at the true value 
$\psi_0=\psi(\theta_0)$. Hence 
$\cc_n(\psi_0,\calD_n)=\Gamma_1(\dev_n(\psi_0))\arr_d\unif$, 
with $\Gamma_1(\cdot)$ denoting the $\chi^2_1$ 
distribution function, and 
\beq
\label{eq:CDapprox3}
\cc_n(\psi,\calD_{n,\obs})=\Gamma_1(\dev_n(\psi)) 
\eeq 
is our confidence curve. It can reflect asymmetry 
and also likelihood multimodality in the underlying distributions, 
unlike the simpler method of (\ref{eq:CDapprox1}). 
Since a confidence curve can be derived from a proper CD,
via (\ref{eq:cases}), but not always the other way around,
the confidence curve is arguably a more fundamental
notion or concept than a CD. 

There is an extensive literature in probability theory 
and statistics regarding the many ways of fine-tuning 
the distributional approximations associated with 
the first-order normality result (\ref{eq:CDapprox1}) 
and the Wilks theorem for (\ref{eq:deviance}). 
Key words for such methods include Bartletting, 
expansions, modified profiles, saddlepointing, 
bootstrap refinements, prepivoting, etc.; see 
e.g.~\citet{BrazzaleDavisonReid07, BrazzaleDavison08, BNCox94}.  
Many of these methods may then be worked with further 
to yield fine-tuning instruments for CDs and confidence curves.
Some of these translations, from the more traditional
setup of assessing accuracy of a certain approximation,
or how to correct for a type of bias, are fairly 
straightforward, leading to good CD recipes. 
Other such translations, involving perhaps higher-level 
bootstrapping or modified log-likelihood operations, 
are non-trivial. Interestingly, some of the more intricate
procedures, like Barndorff-Nielsen's `magic formula',
have relatively speaking easier cousins in the CD 
universe of things, and potentially with easier explanations; 
see \citet[Ch.~7]{CLP16} for discussion and illustrations. 

\begin{figure}[h] 
\begin{center}
\includegraphics[scale=0.66]{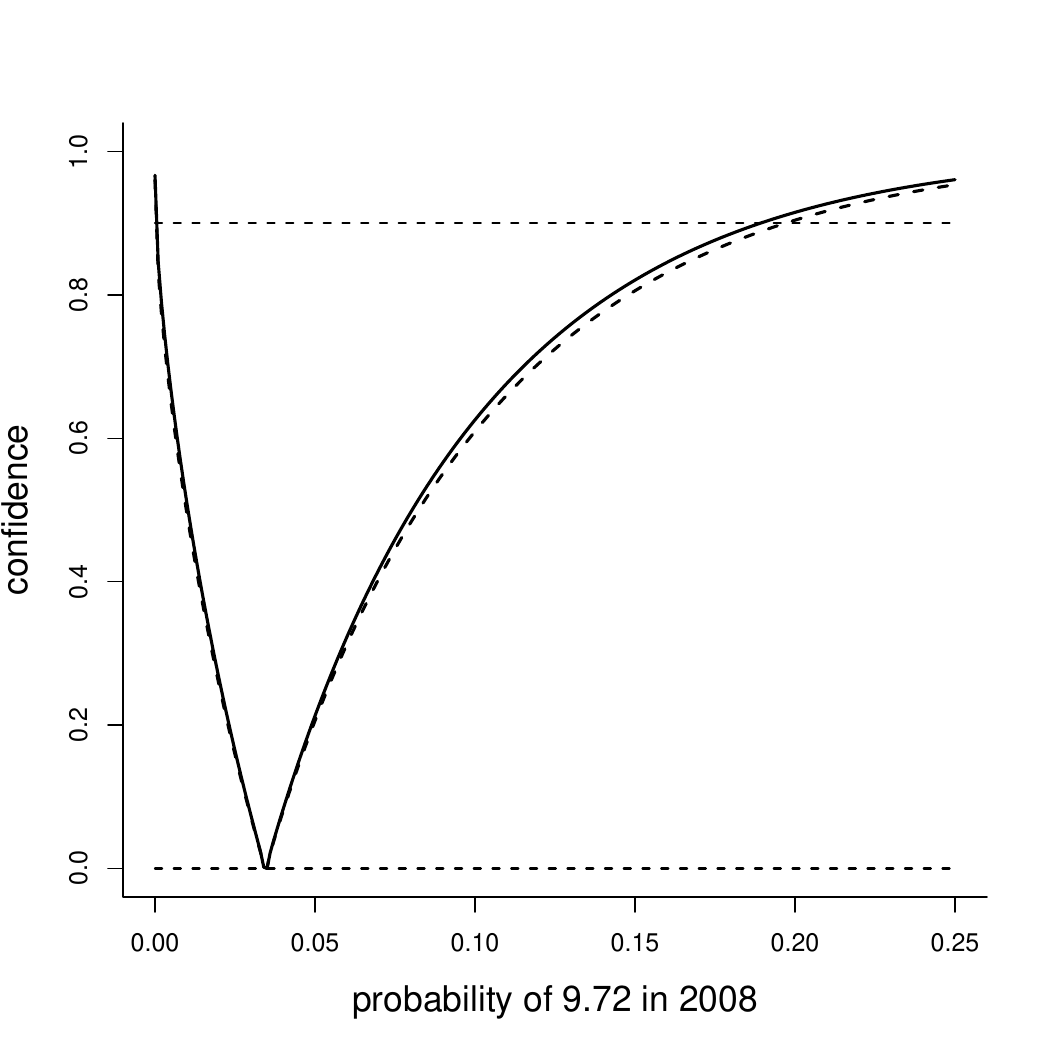}
\end{center}
\caption{Confidence curve for the probability $p$
that there would be a 100 m race of 9.72 or better,
in the course of 2008, as seen from January~1 that year.
The point estimate is 0.034, and the 90\% confidence 
interval is $[0,0.189]$. The dotted curve is a fine-tuned 
version of (\ref{eq:CDapprox3}), via Bartletting.} 
\label{figure:bolt}
\end{figure}

A confidence curve analysis is often much more
informative than providing the prototypical 95\%
interval or a p-value for an associated hypothesis test. 
Figure \ref{figure:bolt} displays the confidence curve 
$\cc(p)$ for the probability $p$ that the world would see 
a 100 m sprint race in a time of 9.72 seconds or faster,
inside the calendar year 2008, with this question
asked on January~1 that year. In other words, this
is an attempt to quantify how surprised we ought
to have been, when we learned that Usain Bolt had 
set his first world record, in May that year. 
We have used the general apparatus of extreme value theory
to make such a question precise, taking as data
the $n=195$ races (which we were able to track down from
various sources) with a result time of 10.00 or better,
in the course of the eight calendar years 2000--2007.
Theory for extreme values leads to a certain 
parametric form for the best races, involving 
parameters $(a,\sigma)$ (and the model has been shown 
to fit very well to the sprint data). The $\cc(p)$ given 
in the figure has come about by (i) expressing 
$p$ as a function of $(a,\sigma)$, (ii) using 
the log-likelihood function $\ell_n(a,\sigma)$
to arrive at the profile and deviance function 
for $p$; and (iii) applying (\ref{eq:CDapprox3}). 
The point estimate is $\hatt p=0.034$, and a 90\% confidence
interval, read off from the figure, is $[0,0.189]$. 
The natural skewness of the distributions involved
makes this a more appealing method than applying 
the traditional $\hatt p\pm1.645\,\hatt\kappa/\rootn$, say. 
The dotted line in Figure \ref{figure:bolt} 
is what here comes out of using a fine-tuning version
of (\ref{eq:CDapprox3}), namely 
$\Gamma_1(\dev_n(p)/(1+\hatt\eps))$, with $1+\hatt\eps$
indicating a Bartlett correction for the distribution
of $\dev_n(p)$. In this particular case, $1+\hatt\eps=1.070$,
and the curves are nearly identical. 
For a fuller discussion and the required detail, 
see \citet[Section 7.4]{CLP16}; see also 
\citet[this issue]{DeblasiSchweder17}, where a novel correction method
for fine-tuning of CDs is applied for this Bolt 2008 problem.  

\section{Risk, performance, optimality, and testing}
\label{section:performance}

Different ways of setting confidence intervals
for the same parameter, and indeed more generally CDs, 
entail different performances. What is reasonably 
to be understood by `good performance',
for a confidence interval or a CD, is less clear 
than for point estimates or tests, where we are 
used to assessing root mean squared errors and 
power curves. Natural classes of loss functions may be
put forward, with the risk functions as usual defined
as the expected values of these losses, as a function
of the the position in the parameter space. Such
themes are developed in \citet[Chs.~5, 7, 8]{CLP16}.
This development may be seen as a natural extension
of classical optimality theory, for testing and 
for point estimation, as with the body of literature 
on Neyman--Pearson testing, etc.; 
see e.g.~\citet{Lehmann59, LehmannRomano05}. 

Here we are content to quote and then illustrate 
a certain optimality theorem, which in particular can be put to use
in models of the classical exponential structure. 
Suppose $\psi$ is a focus parameter, and that 
the log-likelihood function for data can be 
expressed in the form 
$$\ell(\psi,\lambda_1,\ldots,\lambda_k)
   =B\psi+\sum_{j=1}^k A_j\lambda_j-c(\psi,\lambda_1,\ldots,\lambda_k)
   +h(\calD), $$
with nuisance parameters $\lambda_1,\ldots,\lambda_k$,
with $B$ and $A_1,\ldots,A_k$ functions of the data $\calD$,
and appropriate functions $c(\cdot)$ and $h(\cdot)$. 
In that case, the CD 
\beq
\label{eq:Cstar}
C^*(\psi,\calD)=\Pr_\psi\{B\ge B_\obs\midd A_1=A_{1,\obs},\ldots,
   A_k=A_{k,\obs}\}
\eeq 
enjoys optimality properties with respect to a large class
of loss functions for CDs; see \citet[Ch.~5]{CLP16}.  
That this $C^*(\psi)$ depends only on $\psi$, and not 
on the nuisance parameters, is part of the associated
theorems. 

\begin{table}[h]
\caption{Lidocaine data: 
Death rates for two groups of acute myocardial 
infarction patients, in six independent studies,
with control group associated with $(m_0,y_0)$
and lidocaine treatment group with $(m_1,y_1)$; 
from \citet{Normand99}. See Figure \ref{figure:lido}. 
\label{table:lido}} 
\tiny
\begin{center}
\begin{tabular}{ccccc}  
 $m_1$ & $m_0$ & $y_1$ & $y_0$ & $z$\\ 
 39 &  43 &  2 &  1 & 3 \\
 44 &  44 &  4 &  4 & 8 \\
107 & 110 &  6 &  4 & 10 \\
103 & 100 &  7 &  5 & 12 \\
110 & 106 &  7 &  3 & 10 \\
154 & 146 & 11 &  4 & 15 \\
\end{tabular}
\end{center}
\end{table}

\vspace{-0.33cm}
To illustrate this, consider Table \ref{table:lido},
summarising the number of deaths $y_0$ and $y_1$,
with underlying sample sizes $m_0$ and $m_1$,  
in $k=6$ independent studies, involving  
acute myocardial infarction patients. Patients in the 
treatment group, associated with $(m_1,y_1)$, received 
the drug lidocaine; the control group, listed under 
$(m_0,y_0)$, did not; see \citet{Normand99}. 
These are binomial studies, and modelling and analysis
may proceed as in \citet[Ch.~14.6]{CLP16}.
Since the probabilities are small, we choose 
a Poisson model for the present illustration. Our model takes 
\beq
\label{eq:poissonmodel}
y_{j,0}\sim\pois(e_{j,0}\lambda_{j,0}) 
   \hbox{\rm\ and\ } 
y_{j,1}\sim\pois(e_{j,1}\lambda_{j,1}), 
   \quad \hbox{\rm with} \quad 
   \lambda_{j,1}=\gamma\lambda_{j,0},
\eeq   
with exposure numbers $e_{j,0}$ and $e_{j,1}$
proportional to sample sizes $m_{j,0}$ and $m_{j,1}$,
for $j=1,\ldots,k$. Interest focuses on $\gamma$, 
which signals whether the drug use for these patients 
led to an increased death risk. The log-likelihood
for study $j$ takes the form 
\beqn
\ell_j&=&-e_{j,0}\lambda_{j,0}+y_{j,0}\log\lambda_{j,0}
   -e_{j,1}\lambda_{j,0}\gamma+y_{j,1}(\log\lambda_{j,0}+\log\gamma) \\
&=&y_{j,1}\log\gamma+z_j\log\lambda_{j,0}
   -e_{j,0}\lambda_{j,0}-e_{j,1}\lambda_{j,0}\gamma,
\eeqn 
with $z_j=y_{j,0}+y_{j,1}$. The optimality theorem applies, 
involving the distribution of $y_{j,1}\midd z_j$, which
is seen to be a binomial $(z_j,e_{j,1}\gamma/(e_{j,0}+e_{j,1}\gamma))$. 
The optimal CD for $\gamma$, based on study $j$ alone, is hence 
$$C_j^*(\gamma,\calD_j)=1-B(y_{j,1};z_j,e_{j,1}\gamma/(e_{j,0}+e_{j,1}\gamma))
   +\half\,b(y_{j,1};z_j,e_{j,1}\gamma/(e_{j,0}+e_{j,1}\gamma)), $$
with $\calD_j$ signifying the data from source $j$, and 
with $B(\cdot;n,p)$ and $b(\cdot;n,p)$ denoting the cumulative
and point distribution of a binomial $(n,p)$. Here we are 
using the beneficial half-correction for discreteness,
cf.~\citet[Ch.~3.7]{CLP16}. 

\begin{figure}[h] 
\begin{center}
\includegraphics[scale=0.44]{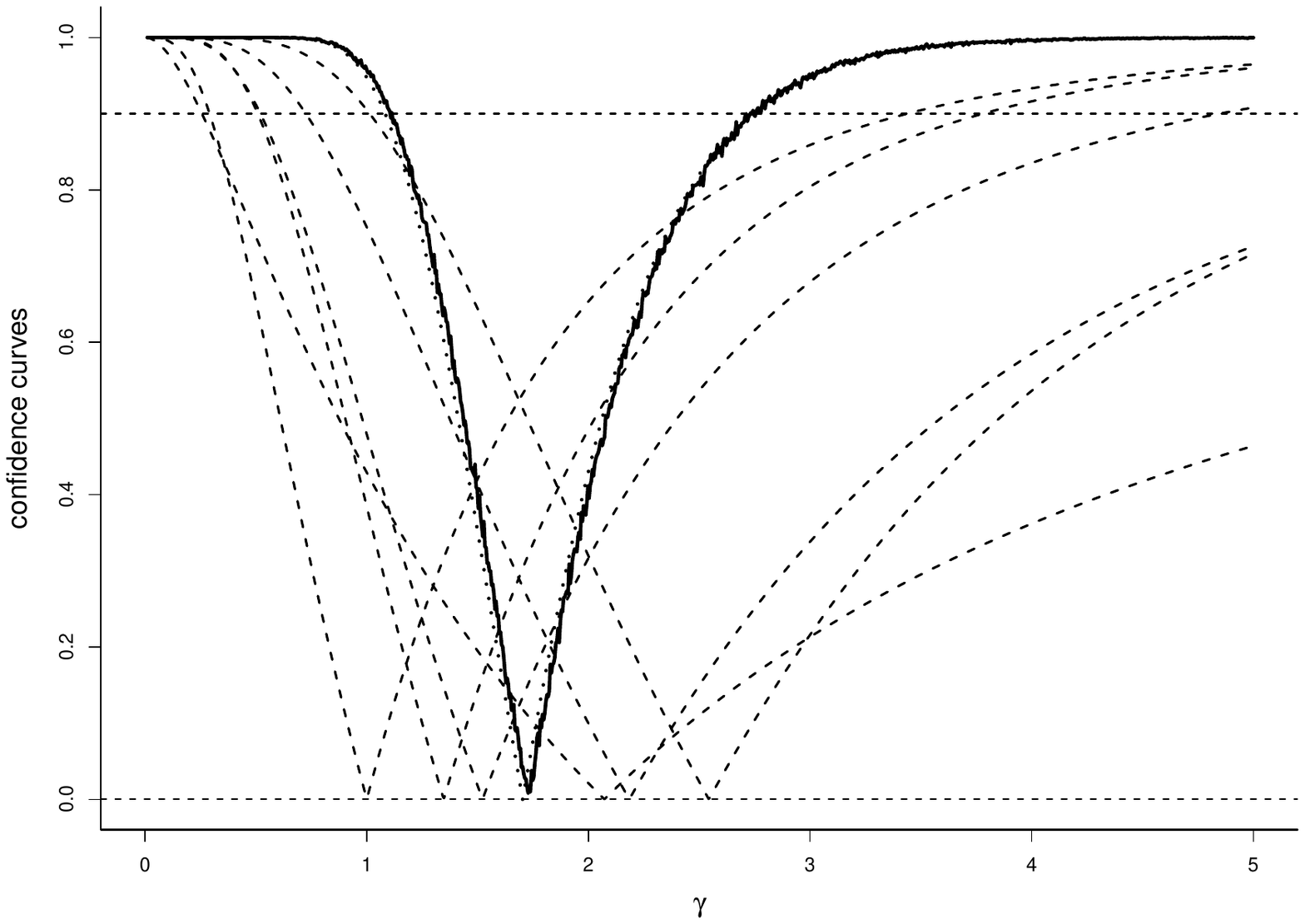}
\end{center}
\caption{The dashed lines are the confidence curves 
for the risk inflation parameter $\gamma$ from each of the six 
studies, from the model (\ref{eq:poissonmodel}) with 
the lidocaine data of Table \ref{table:lido}. 
The thick black curve is the optimal combined confidence curve, 
while the virtually identical dashed curve is the combined 
confidence curve based on the II-CC-FF methods of 
Section 5, without using the Poisson model properties per se.}
\label{figure:lido}
\end{figure}

The $k=6$ confidence curves 
$\cc_j^*(\gamma,\calD_j)=|1-2\,C^*_j(\gamma,\calD_j)|$
for the risk inflation parameter $\gamma$ coming out of this
are seen in Figure \ref{figure:lido} (the dashed curves). 
Also displayed is the overall optimal confidence curve for $\gamma$
(the fatter, full curve), emerging from studying the combined log-likelihood,
$$\ell=\sum_{j=1}^k\ell_j
   =B\log\gamma+\sum_{j=1}^k z_j\log\lambda_{j,0}
   -\sum_{j=1}^k (e_{j,0}+e_{j,1}\gamma)\lambda_{j,0}, $$
with $B=\sum_{j=1}^k y_{j,1}$, and where our optimality theorem leads to 
\beq
\label{eq:Cstargamma}
\begin{array}{rcl}
C^*(\gamma,\calD)&=&\Pr_\gamma\{B>B_\obs
   \midd z_1=z_{1,\obs},\ldots,z_k=z_{k,\obs}\} \\
  &&\qquad   +\half\,\Pr_\gamma\{B=B_\obs
   \midd z_1=z_{1,\obs},\ldots,z_k=z_{k,\obs}\}, 
\end{array}
\eeq 
with $\calD$ denoting the full dataset. 
This is evaluated numerically by simulating a large enough number
of $B$, for each $\gamma$ on a grid of such values, 
from the distribution of a sum of $k$ binomials with
different sets of parameters. 


The main interest for the analysis of the lidocaine dataset 
is the assessment of the risk inflation, if present, i.e.~the
degree to which the treatment for these patients 
leads to increased risk of death. In our Poisson model 
(\ref{eq:poissonmodel}), this is measured via 
the parameter $\gamma$. 
The perhaps most traditional statistical approach
is to test the null hypothesis $H_0\colon\gamma\le1$
versus the alternative that $\gamma>1$. 
As Figure \ref{figure:lido} reveals, there is often 
more information in conducting a full confidence curve 
analysis than in executing a test with its traditional
yes-or-no answer at a certain level of significance, 
like the ubiquitous 0.05. The $\cc^*(\gamma,\calD)$
reveals not merely the overall point estimate 1.732,
but the 0.95 interval $[1.023,3.027]$, along with 
all other intervals; also, the configram clearly reveals
the relative influence of each of the $k=6$ 
separate information sources. The p-value can also 
be read off, from $p=C^*(1,\calD)$, the epistemic
confidence that $\gamma\le1$; the value is 0.021. 

\begin{figure}[h] 
\begin{center}
\includegraphics[scale=0.66]{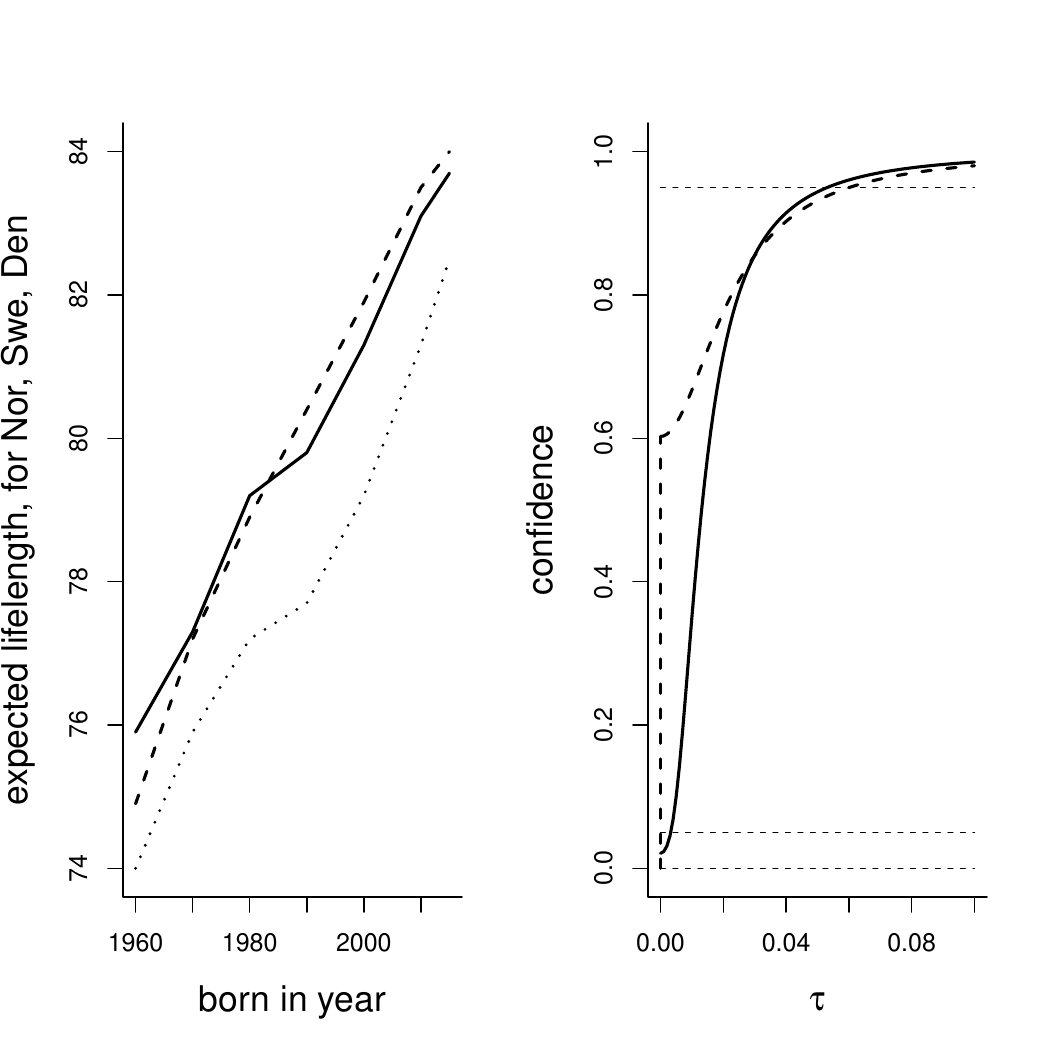}
\end{center}
\caption{Left panel: The expected mean lifelength for 
women born in Norway (full curve), Sweden (dashed curve),
Denmark (dotted curve), in calendar years 
1960, 1970, 1980, 1990, 200, 2010, 2015. 
Right panel: The CD $C(\tau,\calD)$ for the spread parameter 
in the model $\beta_1,\beta_2,\beta_3\sim\N(\beta_0,\tau^2)$
for the three regression slope parameters; 
for men (dashed curve, starting at 0.603 at zero) 
and for women (full curve, starting at 0.021 at zero).
95\% intervals for $\tau$ are $[0,0.060]$ for men
and $[0.003,0.053]$ for women.}
\label{figure:demography}
\end{figure}

As another illustration of this general point, about
how CD analyses and plots often convey more statistical 
information than simple accept-or-reject answers
from carrying out a test, consider 
Figure \ref{figure:demography}, with the left panel
showing the increase in expected lifelength for women born
in Norway (full curve), Sweden (dashed curve), Denmark
(dotted curve), for the years 
1960, 1970, 1980, 1990, 2000, 2010, 2015,
from the website 
{\tt worldlifeexpectancy.com/history-of-life-expectancy}. 
The growth in expected lifelength is amazingly linear,
for this span of calendar time, and we view the 
data as three linear regressions,
say $y_{i,j}=\alpha_i+\beta_i x_j+\eps_{i,j}$ 
for countries $i=1,2,3$ and calendar years $x_j$ 
represented by $j=1,\ldots,7$, and with error terms 
modelled as independent and $\eps_{i,j}\sim\N(0,\sigma_i^2)$. 
We may query whether the regression
slope coefficient $\beta$ is the same for the three
Scandinavian countries. Rather than merely testing the 
hypothesis $H_0$ that $\beta_1=\beta_2=\beta_3$,
which would be standard (the point estimates are 
0.140, 0.162, 0.144, with considerable overlap in
their 0.95 confidence intervals), we address the question 
by modelling these three $\beta$ coefficients as coming
from a background $\N(\beta_0,\tau^2)$ model;
hence $H_0$ is the same as $\tau=0$. Using methods of 
\citet[Ch.~13]{CLP16}, we can derive and compute 
full CDs $C(\tau,\calD)$ for the spread parameter, displayed 
in the right panel. For the three male regressions 
(not shown here), the CD has a big point-mass 0.603 at $\tau=0$; 
there is hence no reason to reject $H_0$, and 
confidence intervals at all reasonable levels start at zero 
(a 90\% interval is $[0,C^{-1}(0.90,\calD)]=[0,0.060]$). 
For the female regressions, however, there are noticeable 
differences in the three slopes underlying what is seen 
in the left panel; the p-value is $C(0,\calD)=0.021$, 
and a 90\% interval is 
$[C^{-1}(0.05,\calD),C^{-1}(0.95,\calD)]=[0.003,0.053]$. 

\section{Data fusion via CDs}
\label{section:fusion}


Meta-analysis is a well-developed area of theoretical
and applied statistics, having to do with the 
comparison, assessment and perhaps ranking of different 
parameters across similar studies. Typical applications 
include analyses of different schools, or hospitals, 
or sport teams, or departments of statistics. Over the past few years 
these topics and methods have been expanded further,
to account for the need to fuse together information
from potentially very different types of sources,
also in connection with the Data Science exploitation of Big Data. 
It is also important in various application areas
to combine Bayesian with frequentist information,
as discussed in \citet{LiuLiuXie15} and \citet[this issue]{Lewis17};
also, \citet[this issue]{Gruenwald17} touches on 
ways in which to handle multiple priors.  

Suppose in general that data source $y_j$ carries information
about parameter $\psi_j$, for sources $j=1,\ldots,k$. 
We wish to assess overall aspects of these $\psi_j$, 
perhaps aiming for inference concerning 
one of more functions $\phi(\psi_1,\ldots,\psi_k)$. 
Let us first assume that the $\psi_j$ parameter 
is the same, across studies, and that the separate 
studies have led to CDs $C_j(\psi,y_j)$. 
A class of methods for combining these is as follows; 
see \citet{SinghXieStrawderman05, XieSingh13, LiuLiuXie14}
and further references therein.  
Under the true value, $C_j(\psi,Y_j)\sim\unif$, 
from which follows $\Phi^{-1}(C_j(\psi,Y_j))\sim\N(0,1)$. 
With weights $w_j$ nonrandom and satisfying $\sum_{j=1}^k w_j^2=1$,
therefore, 
$$\bar C(\psi,\calD)=\Phi\Bigl(\sum_{j=1}^k 
   w_j\Phi^{-1}(C_j(\psi,Y_j))\Bigr), $$
with $\calD$ the full dataset, 
is a CD for the common interest parameter $\psi$. 
Other start ingredients than the normal could also be put
to use, but with less amenable convolutions and inversions. 
This is a versatile and broadly applicable method, 
but with some drawbacks. There are difficulties when 
estimated weights $\hatt w_j$ are used, and there is 
lack of full efficiency. In various cases, there are better 
CD combination methods, with higher confidence power; 
see the discussion in \citet{CunenHjort16}. 


In clearly structured cases, as with several of the 
simpler meta-analysis setups, one can work with the 
full likelihood of the observed data, and deduce good
CDs for interest parameters, see \citet[Ch.~13]{CLP16}. 
This does sometimes require the full set of raw data, 
however, which is often a too tall order. General ways 
of dealing with data fusion with CDs are discussed 
and applied in \citet{LiuLiuXie15} and \citet[this issue]{Hannigetal17}.
Here we describe a more general setup for carrying out data fusion, 
via CDs, which we call the II-CC-FF paradigm; 
see \citet{CunenHjort16}. It is a more 
broadly applicable formulation of likelihood synthesis
ideas first proposed, developed and applied in 
\citet{SchwederHjort96, SchwederHjort97}, in the 
specific context of population dynamics models 
for whale abundance. 

\begin{itemize}
\item[II] {\it Independent Inspection:} 
   From data source $y_j$ to estimate and confidence analysis, 
   yielding a CD $C_j(\psi_j,y_j)$; 
   $y_j\Longrightarrow C_j(\psi_j,y_j)$. 
\item[CC] {\it Confidence Conversion:} 
   From the CD to a confidence log-likelihood, $\ell_{c,j}(\psi_j)$; 
   $C_j(\psi_j,y_j)\Longrightarrow \ell_{c,j}(\psi_j)$.  
\item[FF] {\it Focused Fusion:} 
   Using the combined confidence log-likelihood 
   $\ell_c=\sum_{j=1}^k \ell_{c,j}(\psi_j)$
   to construct a CD for the given focus $\phi=\phi(\psi_1,\ldots,\psi_k)$,
   perhaps via profiling, median-Bartletting, etc.; 
   $\ell_c(\psi_1,\ldots,\psi_k)\Longrightarrow\bar C_{{\rm fusion}}(\phi,\calD)$,
   with $\calD$ denoting the combined dataset. 
\end{itemize}

The FF step, which may also be described as the 
Summary of Summaries operation, will typically involve 
log-likelihood profiling and operations like 
(\ref{eq:CDapprox3}), perhaps along with fine-tuning operations
for increased accuracy. 
Sometimes the CC step is the more difficult one, since 
a clear translation from confidence to likelihood 
often would involve details of sampling design and protocol,
etc. Under mild conditions, however, the {\it normal conversion} 
works well, which is  
$$\ell_{c,j}(\psi_j)=-\half\,\Gamma_1^{-1}(\cc_j(\psi_j,y_j))
   =-\half\{\Phi^{-1}(C_j(\psi_j,y_j))\}^2, $$
cf.~(\ref{eq:CDapprox3}). 

For an illustration, let us go back to the lidocaine 
story of Table \ref{table:lido} and Figure \ref{figure:lido},
for which we have already displayed the optimal 
meta-analysis confidence curve (\ref{eq:Cstargamma}) 
for the risk inflation parameter $\gamma$. We may 
however attempt the II-CC-FF recipe, which leads to a 
$\bar C_{{\rm fusion}}(\gamma,\calD)$ just from converting 
the $k=6$ individual $\cc_j^*(\gamma,\calD_j)$ curves, using
normal conversion, but without using the raw data per se,
or any further knowledge of the underlying Poisson
nature details of the modelling of the data. 
Amazingly, this FF fusion curve is almost indistinguishable
from the $C^*(\gamma,\calD)$. 
 
\section{CDs in semi- and nonparametric situations}
\label{section:nonpara}

The CDs and confidence curves may also be constructed 
in non- and semiparametric situations. By arguments above,
as long as there is an estimator $\hatt\psi$ for 
the required interest parameter $\psi$, with an 
associated limit distribution (typically normal), 
we may construct a CD for $\psi$ based on that estimator. 
The empirical likelihood may also be worked with 
to produce nonparametric CDs, in broad classes of situations, 
as developed and illustrated in \citet[Ch.~11]{CLP16}. 

\begin{figure}[h] 
\begin{center}
\includegraphics[scale=0.6]{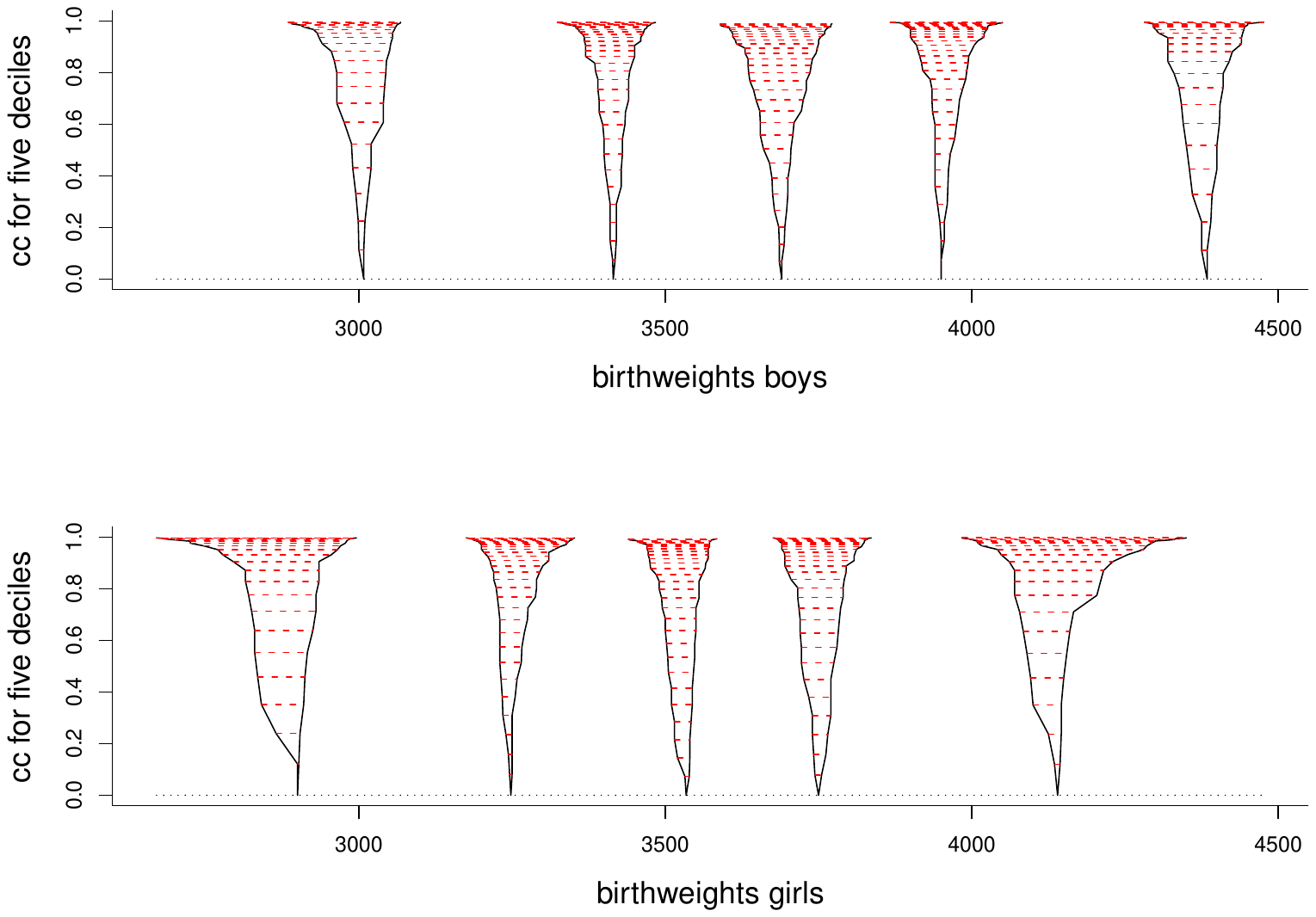}
\end{center}
\caption{Confidence curves $\cc(q)$ for deciles 
0.1, 0.3, 0.5, 0.7, 0.9 of birthweight distributions, 
for boys ($n=548$) and girls ($n=480$) 
born in Oslo 2001--2008.}
\label{figure:deciles}
\end{figure}

In some cases a more exact analysis is possible. 
A case in point is the following, where inference 
is required for the quantiles $\mu_p=F^{-1}(p)$ 
of a continuous and increasing distribution function,
based on i.i.d.~data $y_1,\ldots,y_n$. From the 
fact that the vector of ordered observations $y_{(i)}$ 
has the same distribution as that of $F^{-1}(u_{(i)})$,
where the $u_{(i)}$ are the ordered sample from a 
uniform distribution, we can compute 
$$s_n(a,b)=\Pr\{Y_{(a)}\le\mu_p\le Y_{(b)}\}
   =\Pr\{U_{(a)}\le p\le U_{(b)}\} $$
for each pair $(a,b)$; see \citet[Ch.~11]{CLP16}. 
This can then be used to compute and display confidence curves
$\cc(\mu_p,y)$ for each $p$ of interest, as a nested
sequence of confidence intervals. 
An illustration is given in Figure \ref{figure:deciles},
where we give the full confidence curves for 
the 0.1, 0.3, 0.5, 0.7, 0.9 deciles for the 
birthweight distributions of boys and girls,
born in Oslo, 2001--2008. The $\cc(\mu_p,y)$ curves
tend to be slimmer where there is more data, 
i.e.~around the median on this occasion.

In nonparametric situations there are often parameters
which cannot be estimated at the usual $\rootn$ rate. 
\citet{KimPollard90} give an overview of classes of cases
for which the estimator $\hatt\psi$ for the focus parameter 
$\psi$ in question exhibits cube-root convergence 
in distribution, i.e.~$n^{1/3}(\hatt\psi-\psi)\arr_d L$
for the appropriate (and non-normal) limit $L$. 
With appropriate extra efforts, involving the 
limit distribution and a consistent estimator for its variance, 
say $\hatt\tau$, one may construct CDs of the type 
$K(n^{1/3}(\psi-\hatt\psi)/\hatt\tau)$, perhaps along
with further fine-tuning.   


\section{Robust CDs for parametric models} 
\label{section:robust}

The standard theory for parametric models evolves around 
the use of likelihood methods. This is also at least partly 
the case for the theory and applications of CDs and confidence curves
\citep{XieSingh13, CLP16}. The basic concepts and recipes 
are however not limited to likelihoods per se, 
and various robust alternatives may be worked with. 
To illustrate such general ideas and tools, suppose 
independent observations $y_1,\ldots,y_n$ stem
from an unknown density $g$, and that one wishes to 
fit the data to a parametric model, say $f_\theta=f(\cdot,\theta)$, 
with $\theta$ a $p$-dimensional parameter. Consider 
$$d_a(g,f_\theta)=\int\{f_\theta^{1+a}-(1+1/a)gf_\theta^a
   +(1/a)g^{1+a}\}\,\dd y, $$
for a positive tuning parameter $a$. This is a divergence
(nonnegative, and zero only if $g=f_\theta$), and 
for $a\arr0$ one finds the Kullback--Leibler
divergence $\int g\log(g/f_\theta)\,\dd y$ associated
with the maximum likelihood method. The BHHJ method,
from \citet{BHHJ98, JHHB01}, estimates $\theta$ by
minimising an empirical version of $d_a$, namely 
$H_n(\theta)=\int f_\theta^{1+a}\,\dd y
   -(1+a/n)\,n^{-1}\sumin f(y_i,\theta)^a$. Setting 
the derivatives equal to zero, the BHHJ estimator is 
also the solution to the equations 
$$n^{-1}\sumin f(y_i,\theta)^a u(y_i,\theta)=\int f_\theta^{1+a}u_\theta\,\dd y, $$
where $u_\theta(y)=u(y,\theta)=\dell\log f(y,\theta)/\dell\theta$
is the score function for the model. Contributions 
from data points with low probability under the model 
thus get weighted down. The method is a successful
robustification of the maximum likelihood strategy
(also in regression setups and other models more 
elaborate than the i.i.d.~situation considered here), 
earning bounded influence functions at the expense
of a very mild loss of efficiency under perfect model
conditions, if $a$ is small.  

\begin{figure}[h] 
\begin{center}
\includegraphics[scale=0.66]{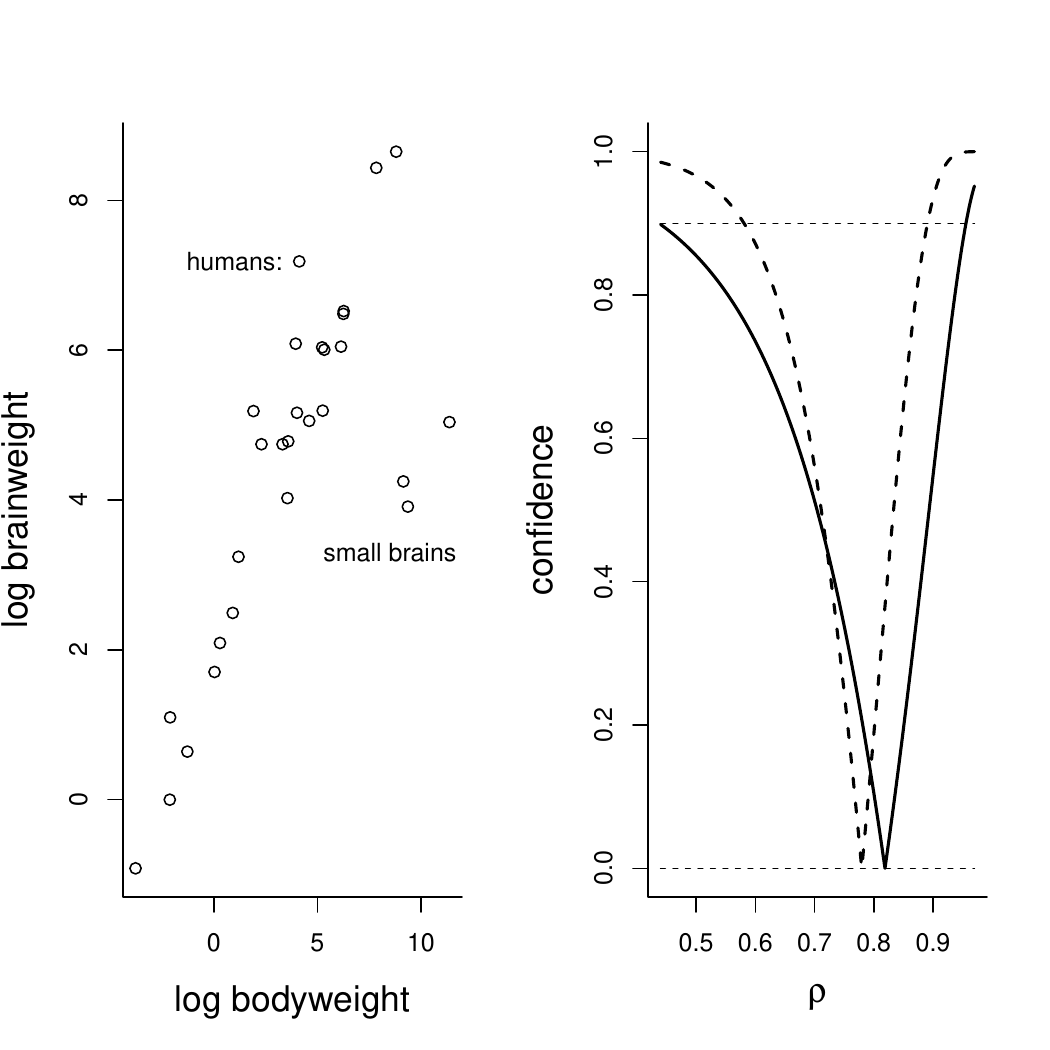}
\end{center}
\caption{Left panel: with average $x_0$ bodyweight (in kg) and 
average brainweight $y_0$ (in g), for 28 species of 
land animals, the plot gives $(x,y)=(\log x_0,\log y_0)$. 
Right panel: two confidence curves for the correlation
coefficient $\rho$, based on maximum likelihood 
(estimate 0.779) and one using the robust BHHJ method 
(estimate 0.819).} 
\label{figure:animals}
\end{figure}

The present point we wish to make is that the criterion function
$H_n$, used to find the BHHJ estimator and its approximate
multinormal distribution, can also be profiled, leading
to confidence curves for focus parameters. With 
$\psi=\psi(\theta)$ such a focus parameter, the 
BHHJ estimator is $\hatt\psi=\psi(\hatt\theta)$, and we form 
$H_{n,\prof}(\psi)=\min\{H_n(\theta)\colon \psi(\theta)=\psi\}$ and then 
the associated deviance function, 
$$D_n(\psi)=2n\{H_{n,\prof}(\psi)-H_{n,\prof}(\hatt\psi)\}
   =2n\{H_{n,\prof}(\psi)-H_{n,\min}\}. $$
With arguments along the lines of 
\citet[Ch.~2.4, Appendix A.6]{CLP16}, one may establish
that $D_n(\psi_0)\arr_d k\chi^2_1$, 
at the appropriate least false parameter value 
$\psi_0=\psi(\theta_0)$, with $\theta_0$ minimising 
the distance $d_a(g,f_\theta)$ from the true $g$ to 
the parametric model. Here $k$ is a certain extra factor
which may be estimated consistently from the data.
This leads to the robust confidence curve 
$\cc(\psi,\calD_n)=\Gamma_1(D_n(\psi)/\hatt k)$ 
(again with $\calD_n$ denoting the dataset), 
in generalisation of (\ref{eq:CDapprox3}).   

This machinery works also for multidimensional data.
Figure \ref{figure:animals} relates to an illustration
of this, where we have studied the dataset {\it Animals}
in R, with $(x_0,y_0)$ equal to average bodyweight (in kg)
and average brainweight (in g) for $n=28$ species 
of land animals. On the log-and-log scale of 
$(x,y)=(\log x_0,\log y_0)$, intriguingly, the points 
nearly form a linear regression structure; the deviants,
from this perspective, are the big-brained humans,
and the big-bodied small-brained Brachiosaurus, 
Triceratops, and Diplodocus (left panel).  
Our chosen focus, for this illustration, is the correlation
coefficient $\rho$. The estimate is 0.779, based on all 28
species, but a much higher 0.960 if we remove the three
small-brained just mentioned. We fit the five-parametric
binormal model to the data, first using maximum likelihood
analysis, then the BHHJ method with $a=0.105$; this value 
makes data pairs an average distance away from the centre,
as measured by the Mahalanobis distance, be downweighted
10\% (and pairs further away from the centre will be 
downweighted more). This value also ensures good robustness.
The two confidence curves are displayed in the right panel;
the maximum likelihood version points to $\hatt\rho=0.779$
whereas the BHHJ method has $\hatt\rho_a=0.819$. 
The robust 90\% confidence interval is $[0.441,0.955]$.
Importantly, these two confidence curves do not 
assume that the binormal model holds. 
In this particular application the robust BHHJ method 
leads to a somewhat broad confidence curve, since 
the method attempts to fit a somewhat non-homogeneous 
dataset to a single binormal density. 
For larger values of $a$, the BHHJ estimation method
will indirectly downweight the three outliers more fully,
and the correlation estimate will come closer to 0.960.    

\section{Bayes versus CDs} 
\label{section:bayes}

The Holy Grail of statistics Brad Efron alludes to 
is to enjoy the Bayesian omelet 
without breaking the Bayesian eggs \citep{Efron10}. It was 
the non-existence of a non-informative prior which led Fisher 
to fiducial distributions. That a CD is `posterior' 
without any prior is its main selling points. 

Bayes' formula is of course true, and the Bayesian posterior 
is the correct updating of a trustworthy prior. But problems 
arise when there is no trust in the chosen prior, or when 
there are more than one legitimate priors.

With much data the CD will tend to be close to the Bayesian 
posterior, by various \Bernstein-von Mises type theorems 
(see e.g.~\citet[Introduction]{HHMW10}). This might also 
happen in some cases with moderate and small data, particularly 
when a Jeffreys prior is used. A case of the latter is seen in 
\citet[this issue]{Lewis17}, where he develops
both a Byesian posterior and a CD for the climate sensitivity. 
They are seen to be indistinguishable.

A marginal posterior distribution might be misleading, 
as illustrated by the so called length problem: 
With independent $Y_i\sim\N(\mu_i,1)$ for $i=1,\ldots,m$, 
the marginal posterior for $\psi=\|\mu\|=(\sum_{j=1}^m\mu_j^2)^{1/2}$ 
based on a flat (Jeffreys) prior for the $m$ mean parameters 
is biased in the frequentist sense that its credibility intervals 
will not have correct coverage probabilities 
\citep[Section 9.4]{CLP16}. The distribution 
is actually shifted to the right relative to $\psi$ 
and more so the larger $m$ is. This is also a problem 
for the marginal of Fisher's joint fiducial distribution, 
which is not a CD. A similar bias inherent in Bayes
setups is noted for change-point assessments 
in higher dimensions, in \citet[this issue]{CunenHermansenHjort17}.

Potential bias seems not to be a concern for most Bayesians. 
When your prior is to be updated from new data, you get 
the posterior you get, and performance in repeated applications 
is seen as irrelevant. Frequently the model is complex 
and the model parameter of substantial dimensions, however, 
as in the length problem. As a more realistic example consider 
the parameter $\theta_1$ of interest to \citet{Sims12} 
in his Nobel Memorial Prize in Economic Sciences acceptance lecture,
where it is also argued that $\theta_1\ge0$ on 
a priori grounds. The model is a linear simultaneous 
equations model for macroeconomic data. 
The chosen prior for coefficients, including $\theta_1$, 
is flat. Since the unrestricted
maximum likelihood $\hatt\theta_1$ is negative the posterior 
is shifted to the right of the CD for $\theta_1$. The latter
has actually a point mass of $0.90$ at zero 
(when the restriction is $\theta_1\ge0$), while Sims's  
posterior has all its mass on the positive values; 
see \citet[Section 14.4]{CLP16}. 

Bayesian methods are very often used. It is thus a bit odd 
that performance in repeated applications is mostly neglected.
Bias and other frequentist properties are however of concern 
to some Bayesians. The invariance of the posterior based 
on Jeffreys priors, to transformations of the model parameter, 
will, as noted above, make the posterior nearly or 
exactly a CD. In cases with a parameter $\psi$ of interest, 
of lower dimension than the model parameter, the objective 
Bayesian uses a reference prior \citep{BergerBernardo92, BergerSun08} 
tailored to $\psi$. This is parallel 
to confidence inference where new calculations
are needed for each $\psi$. The posterior based on a reference 
prior aims at having correct coverage probabilities for 
its credibility regions in repeated applications. The CD 
has the same aim -- it is actually its defining property. 
The realised posterior and also the CD are understood as epistemic
probability distributions for $\psi$ 
\citep[this issue]{Schweder17}.  
To be a CD might actually be the gold standard for
an epistemic probability distribution for a parameter of interest, 
at least for the objective Bayesian; 
cf.~\citet[this issue]{VeroneseMelilli17} and also 
\citet[this issue]{Gruenwald17}. 
\citet{Fraser11} actually suggests that Bayes posterior distributions 
are just quick and dirty confidence distributions.

We note that CDs may be constructed not only for 
parameters of models, but also for not-yet-seen 
random variables, as in prediction contexts. There
are again similarities with Bayesian approaches; 
see \citet[Ch.~12]{CLP16} and \citet[this issue]{Shenetal17}. 

An important virtue of the Bayesian approach is its coherence. 
Ordinary probability calculus applies to posterior distributions. 
For CDs some probability calculations yield new CDs, while others
do not. When the prior has been set up, the challenge is to calculate 
the well-defined posterior, say by Markov Chain Monte Carlo. 
The technical virtuosity of current days Bayesians is really 
impressive, and has led to sensible analyses of complex data 
in many areas. The machinery for confidence inference is
by far less developed. Significant applications in science 
are still rather few (but see \citet{CunenLarsNils17}, where 
the main findings were communicated to the Scientific Committee
of the International Whaling Commission via confidence curves).
Software for CDs needs further development in good packages
in order for the dissemination to gain momentum. 

\citet{Robert13} points out that a CD is in essence 
just a representation of a nested family of confidence regions, 
and as such not particularly novel, per se. The emphasis 
on CDs as distributions on par with Bayesian posteriors 
might however be a rather novel insight, distributions that 
``provide simple and interpretable summaries of what 
can reasonably be learned from data (and an assumed model)'' 
\citep{Cox13}. There is also scope for novel and streadily 
more impressive uses of CDs for data fusion, when information 
sources are more diverse than in the typical meta-analysis 
setups; see \citet{XieSingh13, LiuLiuXie15, CunenHjort16}.  



\section{The present collection of papers}
\label{section:collection}

Articles appearing in the present Special Issue
have been mentioned above, in the relevant contexts.
Here we offer just a few more comments to help 
readers navigate through these contributions 
and to see connections between them. 

Each of the contributions \citet{Hannigetal17}, 
\citet{LindqvistTaraldsen17}, \citet{TaraldsenLindqvist17}
deal with fiducial and generalised fiducial inference
questions, also with relevance for the eternal comparison
with Bayesian constructions. Articles 
\citet{DeblasiSchweder17}, \citet{VeroneseMelilli17} 
are partly concerned with fine-tuning mechanisms 
for the constructions of CDs, with further connections
to so-called objective Bayes. Several contributions
are involved with the important topic of combining
information across diverse sources, sometimes called
data fusion: \citet{Hannigetal17}, \citet{Gruenwald17}, 
\citet{CunenHermansenHjort17}, \citet{Lewis17}. 
The latter paper is also a well-argued contribution
to the always hot topic of climate research, where
there typically are very different sources of information.
One of the challenges, worked with by Lewis, is that 
of combining summaries reached by Bayesian and 
frequentist perspectives; see also \citet{LiuLiuXie15} 
and \citet{CunenHjort16}. 

When constructing CDs, and more generally machineries
aspiring to deliver posteriors without priors, one
is often close to the more fundamental issues and ideas
of how probability can or should be defined and interpreted,
cf.~again the half-eternal half-disagreements between
Bayesians and frequentists. This is also touched on 
in the essay \citet{Schweder17}, somewhat indirectly
in \citet{Hannigetal17}, \citet{Martin17},
and by \citet{Gruenwald17}. In certain application areas
it might be natural to interpret confidence in 
the language of epistemic probabilities, as argued 
by \citet{Schweder17}; see in this connection also \citet{Helland18}. 

The \citet{Shenetal17} and \cite{CunenHermansenHjort17} articles 
are occupied with respectively prediction issues, 
e.g.~for time series models, and with estimating and 
assessing change-points and regime-shifts,
in settings with discrete data. Applications in the latter paper 
involve finding when Author B took over for Author A, in the 
world's first ever novel (from 1460), 
and searching for a regime-shift in a complex fisheries model. 
That paper also contains novel goodness-of-fit tests 
for checking whether a probability distribution 
has remained constant over a stretch of time.

\section{Concluding remarks}
\label{section:concluding} 

We started out discussing the Holy Grail of parametric 
inference \citep{Efron10}, that of reaching well-defined 
posteriors for interest parameters without putting 
up priors. We argue that the CDs are the answer, or
part of the answer. In particular, in classes of clear-cut
situations, in exponential class type models where 
the broad optimality theory of \citet[Chs.~5, 7, 8]{CLP16}
applies, the optimal CD provides what a rational statistician
ought to believe about the unknown parameter, given
the model and the data. 

There are of course many remaining issues and obstacles
for our profession to work with and perhaps slowly 
sort out, through the statistical symbiotic machineries 
of good theory and solid practice. Let us mention some of these. 

The serious study of CDs and indeed related themes 
often enough touches the fundamental issues of what 
probability is, or ought to be. This has of course been discussed 
inside and outside academics since around 1665 
(see the engaging account by \citet{Hacking75}), 
and perhaps also the modern statisticians and data scientists
need to accept that there are several valid notions,
living if not always comfortably side-by-side:
The clear aleatory probability; the subjective used 
by strands of Bayesians; the epistemic; and yet further 
cousins and hybrids, like Dempster--Shafer belief functions
\citep{Dempster08, Martin17}. There is still a need for a better 
axiomatic theory for epistemic probability, and its
connections to likelihood theory and related issues.  

On the technical side there is scope for important
work along the lines of further fine-tuning 
of approximate CDs to deliver more accurate coverage,
which in the language of CDs, and of this article,
may be described as calibrating the CD such that 
$C(\psi(\theta_0),Y)$ has a distribution close 
to the unifom, where $\psi=\psi(\theta)$ is the focus
parameter and $\psi_0=\psi(\theta_0)$ is the implied 
true focus value. The basics of a list of relevant techniques
is in \citet[Chs.~7-8]{CLP16}, but there is more 
to do, also in situations with multimodal log-likelihoods,
with growing dimension $p$ compared to sample size $n$, etc.
A class of alternatives to the profiling involved in 
(\ref{eq:CDapprox3}) is via the operation of integrating
out other parameters, which may work well also from
the frequentist viewpoint \citep{BergerLiseoWolpert99}.

Several of the issues that bothered Fisher's contemporaries,
when they hesitated to embrace his fiducial inference
ideas in the 1930ies and 1940ies, have to do with 
the usual probability machinery not being applicable
in general. The distribution 
$H(|\mu|)= C(|\mu|) - C(-|\mu|)$, for example, is usually 
not a CD for $|\mu|$ when $C$ is a CD for $\mu$; 
see \citet{SchwederHjort13} for this and similar examples. 
\citet[Ch.~6]{CLP16} refer to various attempts to figure 
out when a distribution obtained by ordinary probability 
calculus from a CD is itself a CD. These theories 
are far from complete. It is perhaps more fruitful 
to study when good approximate CDs can be obtained 
by ordinary calculus than to try to develop 
a calculus for fiducial distributions and CDs.


An interesting research direction is that of 
matching good CDs, in particular those known to 
be optimal, with corresponding priors, i.e.~for so-called
objective Bayes. This is also touched on in 
\citet[this issue]{VeroneseMelilli17}. There are 
situations, e.g.~those with bounded parameter spaces,
where optimal CDs apparently have no matching prior.
An instance of this is our disagreement with
what \citet{Sims12} claimed in his Bayesian-flavoured 
Nobel Memorial Prize acceptance speech; see 
\citet[Ch.~14.4]{CLP16}. 

An important objection from the Bayesian camp is that CDs 
are usually not well-defined when the model has been 
arrived at via a preliminary model selection step. 
Further decisions are usually needed to guide the calculations. 
\citet{Robert13} calls this `ad hockery'. But is it more ad hockery
than choosing the prior in the absence of solid prior information? 
Also, progress can be expected regarding working out
good refinements for CDs after model selection,
using the machinery developed in \citet{HjortClaeskens03}
and \citet{Hjort14} for frequentist model averaging. 

We have seen how p-values often can be seen as 
special components of a bigger CD picture, and these
connections can be worked out more fully, both
for enhanced interpretation and for better 
assessments of well-understood hypotheses; 
cf.~the still ongoing debate on the uses 
and many misuses of p-values \citep{ASAstatement}.  

The CDs and confidence curves should find more use
in the contemporary world of Big Data, Data Science
and Machine Learning, also for conveying summary information 
about the most pertinent issues, based on often 
complex and massive background data. Instrumental
here is also the task of combining and fusing together
information across very diverse sources, where what
we describe above as the II-CC-FF paradigm ought to
be harnessed further. 

There is already a body of literature and results 
on the performance and optimality of classes of CDs,
cf.~again \citet[Ch.~5]{CLP16}. Aspects of this theory 
ought to be extended from CDs to confidence curves,
as there are natural cases where the $\cc(\psi,y)$
is the more fundamental notion of confidence;
cf.~the Fieller problem, situations with multimodal 
log-likelihoods, etc. There is similarly a need for 
more work and better insights for confidence curves
in higher dimensions. Finally we point to the correlated
worlds of CDs, inferential models and generalised
fiducial inference, where there is a need to sort out 
better when the approaches agree, and where they might not. 


\section*{Acknowledgements}
As guest editors we are first of all grateful for the 
high-quality and patience-demanding efforts of all the authors,
also during the required rounds of revisions and 
cross-referencing. We are also indebted to the Norwegian
Research Council for its funding of the FocuStat 
project (Focus Driven Statistical Inference With Complex Data)  
at the Department of Mathematics, University of Oslo; 
this was also instrumental for the group being able
to host a series of international workshops in Oslo, including 
{\it Inference With Confidence} in May 2015. 
Thanks are in particular due to C\'eline Cunen, 
whose efforts in connection with the full process 
of handling and sharpening the contributions 
to the present Special Issue were crucially helpful, 
along with the help and insights from a select group of 
conscientious referees. We finally acknowledge with
gratitude the active help of Holger Dette and the other editors 
of the JSPI in creating a high-quality special issue. 

\parindent0pt
\baselineskip16pt

\bibliographystyle{biometrika}
\bibliography{jspi-intro.bib}

\end{document}